\documentclass[prd,preprint,tightenlines,showpacs]{revtex4}
\usepackage{amsmath,amssymb,epsfig,capt-of,ifthen,calc}
 
\newcommand{\tmstrong}[1]{\textbf{#1}}
\newcommand{\tmop}[1]{\operatorname{#1}}
\newcommand{\tmfloatcontents}{}
\newlength{\tmfloatwidth}
\newcommand{\tmfloat}[5]{
  \renewcommand{\tmfloatcontents}{#4}
  \setlength{\tmfloatwidth}{\widthof{\tmfloatcontents}+1in}
  \ifthenelse{\equal{#2}{small}}
    {\ifthenelse{\lengthtest{\tmfloatwidth > \linewidth}}
      {\setlength{\tmfloatwidth}{\linewidth}}{}}
    {\setlength{\tmfloatwidth}{\linewidth}}  \begin{minipage}[#1]{\tmfloatwidth}
    \begin{center}
      \tmfloatcontents
      \captionof{#3}{#5}
    \end{center}
  \end{minipage}}
 
\begin{document}

\title{
CP asymmetry in $B \rightarrow \phi K_S$ in a general two-Higgs-doublet
model with fourth-generation quarks }
\author{Yue-Liang Wu}
\affiliation{Institute of theoretical physics, Chinese Academy of Sciences,\\
P.O.Box 2735, Beijng, 100080, P.R.China}
\email[Email:  ]{ylwu@itp.ac.cn}
\author{Yu-Feng Zhou}
\affiliation{Institute for physics, Dortmund University, D-44221, Dortmund, Germany}
\email[Email: ]{zhou@zylon.physik.uni-dortmunde.de}
\begin{abstract}
   We discuss the time-dependent CP asymmetry of  decay $B \rightarrow \phi K_S$  in an
  extension of the Standard Model with both two Higgs doublets and additional
  fourth-generation quarks. We show that although the Standard Model with
  two-Higgs-doublet and the Standard model with fourth generation quarks
  {\tmstrong{alone}} are not likely to largely change the effective $\sin 2
  \beta$ from the decay of $B \rightarrow \phi K_S $, the  model with {\tmstrong{both}}
  additional Higgs doublet and fourth-generation quarks can easily account for
  the possible large negative value of $\sin 2 \beta$ without conflicting with
  other experimental constraints. In this model, additional large CP violating
  effects may arise from the flavor changing Yukawa interactions between
  neutral Higgs bosons and the heavy fourth generation down type quark, which can modify
  the QCD penguin contributions. With the constraints obtained from  $b
  \rightarrow s \bar{s} s$ processes such as $B \rightarrow X_s \gamma$ and
  $\Delta m_{B_s^0}$,  this model can lead to the effective $\sin 2
  \beta$  to be as large as  $- 0.4$ in the CP asymmetry of $B \rightarrow \phi K_S$.
\end{abstract}
 
\preprint{DO-TH 04/02}
\maketitle
\section{introduction}
With the successful running of two $B$ factories in KEK and SLAC,  precise
measurements of the time-dependent CP asymmetries as well as the directly CP
asymmetries in rare $B$ decays become available. Among those interesting
decay modes, the most important one, the CP asymmetry  of  $B \rightarrow J / \psi K_S$ has been
successfully measured, and a very good agreement with the Standard Model (SM)
prediction on  $\sin 2 \beta$  was found.
 
However, the recent Belle results on $\sin 2 \beta$ from $B \rightarrow
\phi K_S$, although with significant errors, have indicated  that
the value of $\sin 2 \beta$ from different decay modes could be
significantly different. The most recent measurements give
{\cite{Abe:2003yt,Aubert:2004ii}}
\begin{eqnarray}
  \sin 2 \beta & = & 0.47 \pm 0.34^{+0.08}_{-0.06} ( \tmop{ Babar} ), \nonumber\\
  \sin 2 \beta & = & - 0.96 \pm 0.5^{+ 0.09}_{- 0.11} ( \tmop{ Belle} ) .
\end{eqnarray}
Of course, it is too early to draw any robust conclusion from the
current preliminary data. Nevertheless, it opens a possibility that
large new physics effects may show up in the $b \rightarrow s s s$
processes, which has already triggered a large amount of theoretical
efforts in examining the possible new physics contributions from
various models. Besides the models related to supersymmetry which are
the most promising ones, there are also a large class of models based
on simple extensions of the matter contents of the SM, such as the
standard models with two-Higgs-doublet (S2HDM)
\cite{Glashow:1977nt,Savage:1991qh,Hou:1992un,Antaramian:1992ya,Hall:1993ca,wu:1994ja,wolfenstein:1994jw,Wu:1994vx,%
atwood:1997vj,Dai:1997vg,Bowser-Chao:1998yp,Diaz:2000cm,Zhou:2000ym,%
Xiao:2002mr,Wu:2001vq,Zhou:2001ew} and the standard model with
fourth-generation fermions
(SM4){\cite{Chou:1984bh,Datta:1989in,Huang:2000xe,Huang:1999dk,Arhrib:2002md}} etc.
However, the most recent studies have pointed out that the
contributions from the above mentioned two types of models to $B
\rightarrow \phi K_S$ are in general not large enough to account
for a large negative  value of $\sin 2 \beta$ in $B \rightarrow
\phi K_S$ ( for example $\sin 2\beta \approx -0.5$ )
{\cite{Hiller:2002ci,Giri:2003jj,Huang:2003bj,Arhrib:2002md}}.
 
In this paper, we show that although due to the constraints from
other experiments such as $b \rightarrow s \gamma$ and $\Delta
m_B$ etc., the general S2HDM  and the SM4 alone are not likely to
largely change the effective $\sin 2 \beta$ in $B \rightarrow \phi
K_S $, a model with {\tmstrong{both}} an additional Higgs doublet
and 4th-generation quarks (denoted by S2HDM4) can significantly
change the  value of $\sin 2 \beta$ without contradicting with
other experimental constraints. In this model, new large CP
violating contributions may arise from the flavor-changing Yukawa
interactions between the neutral Higgs boson and the
4th-generation down type quark $b'$ (with $m_{b'} \gg m_b$), which
changes the Wilson coefficients for QCD penguin operators and
results in a large modification of effective $\sin 2 \beta$. This
mechanism is different from the case in the  S2HDM in which the
dominant contribution comes from changing the Wilson coefficients
of the electro(chromo)-magnetic operators. The latter is subjected
to a rather strong constraint from $b \rightarrow s \gamma$ and
therefore can not give enough contributions.
 
\begin{figure}[htb]
\includegraphics[width=0.8\textwidth]{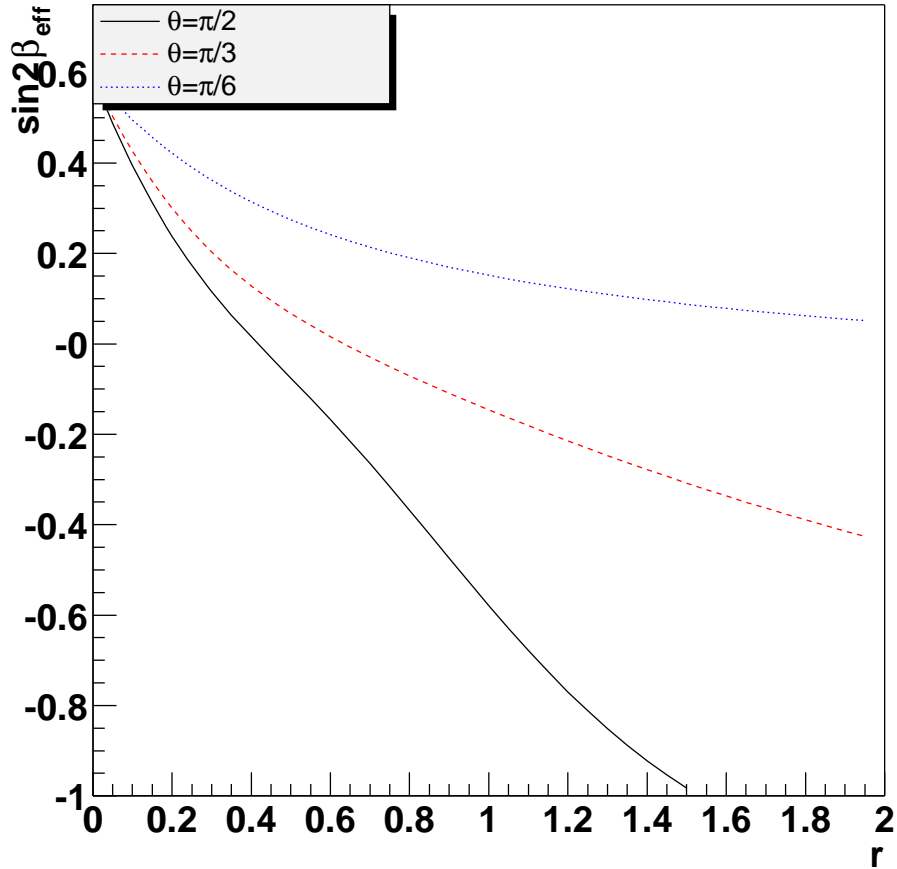}
\caption{The value of $\sin 2 \beta_{\tmop{eff}}$ as a function of
$r$. The solid, dashed and dotted curves corresponds to $\theta = \pi / 2, \pi
/ 3$ and $\pi / 6$ respectively.}
\label{NPCPP}
\end{figure}

Let us begin with some model independent discussions. The definition of
effective $\sin 2 \beta$ in $B\rightarrow \phi K_S$ is
\begin{eqnarray}
  \sin 2 \beta_{\tmop{eff}} & = & \tmop{Im} \left[ e^{2 i \beta}
  \frac{\bar{\mathcal{A}}}{\mathcal{A}} \right] = \tmop{Im} \left[ e^{2 i \beta}
  \frac{\bar{\mathcal{A}}_{\tmop{SM}} ( 1 + r e^{- i \theta}
  )}{\mathcal{A}_{\tmop{SM}} ( 1 + r e^{+ i \theta} )} \right] ,
\end{eqnarray}
where $\beta$ is the SM value  with  $\sin 2 \beta = 0.715^{+ 0.055}_{- 0.045}$
{\cite{Buras:2002sd}}. $\bar{\mathcal{A}}_{\tmop{SM}} (
\mathcal{A}_{\tmop{SM}} )$ is the SM value of the decay amplitude of
$\overline{B^{}}^0 ( B^0 ) \rightarrow \phi K_S$. Here two parameters $r$ and
$\theta$ parameterize the relative size and the additional CP
violating phase of the new physics contributions. To get an idea of
how $\sin 2 \beta_{\tmop{eff}}$ is changes with the new physics
contribution, we take some typical values of the phase $\theta$,
calculate the values of $\sin 2\beta_{\tmop{eff}} $, and shown them  in Fig.\ref{NPCPP}.
 
%
 
As it is shown in the figure, to explain the possibly large negative $\sin 2
\beta_{\tmop{eff}},$ for instance, close to $- 0.5$, in the case that $\theta$ is
maximum ($\pi / 2$), the value of $r$ should be close to unity.  For smaller
$\theta$ such as $\pi/3$ and $\pi/6$, the value of $r$ must be even larger.
Therefore, to generate a large negative value of
$\sin 2 \beta_{eff}$ in the range of $- 0.5 \sim - 1.0$, the magnitude of
the new physics contributions must be as the same order of magnitude as the one in
the SM.
 
However, the new physics contributions must be constrained by
other experiments, especially by the $b \rightarrow s$ transition
related processes. The most strict constraint comes from the
radiative decay of $B \rightarrow X_s \gamma$. The current data of
$\tmop{Br} ( B \rightarrow X_s \gamma ; E_{\gamma} > 1.6
\tmop{GeV} ) = ( 3.28^{+ 0.41}_{- 0.36} ) \times 10^{- 4}$
{\cite{Barate:1998vz,Chen:2001fj,Abe:2001hk,Aubert:2002pd}} is
well reproduced in the frame work of the next-to-leading order
calculations in the SM (see
e.g.\cite{Gambino:2001ew,Buras:2002tp}). Thus, if the new physics
contribution carries no new phase, there is very little room for
the new physics parameters. But in the case that new phases
present, the parameter space could be enlarged. This is because
the data of $B\rightarrow X_s \gamma$ only constraints the absolute values of the  Wilson
coefficient $C_{7 \gamma}$ , if the new physics contribution does
not change the absolute value of $C_{7 \gamma}$, there will not be
a serious problem. Thus the following relation must be satisfied
for any new physics model
\begin{eqnarray}
  \left. \left. \left. \left. \right| C_{7 \gamma} \right| = \left| C_{7
  \gamma}^{\tmop{SM}} + C_{7 \gamma}^{\tmop{NEW}} \right| \simeq \right| C_{7
  \gamma}^{\tmop{SM}} \right|, &  &  \label{c7gamma}
\end{eqnarray}
with $C_{7 \gamma}^{\tmop{SM}}$ and $C_{7 \gamma}^{\tmop{NEW}}$
being the effective Wilson coefficient  evaluated at the low energy scale $(
\mu \approx m_b )$ from SM and new physics models respectively. In
this case, the absolute value of $C_{7 \gamma}^{\tmop{NEW}}$ could
vary largely from close to zero to about $- 2 \text{$C_{7
\gamma}^{\tmop{SM}}$}$, which seems large enough for explaining
the CP asymmetry in $B \rightarrow \phi K_S$. However, it
follows from Eq.(\ref{c7gamma}) that the data on  $B \rightarrow
X_s \gamma$ do strongly constrain the form of $C_{7
\gamma}^{\tmop{NEW}}$, namely, the new physics must interfere in
such a way that the total effect is roughly equivalent to adding a
phase factor to $C_{7 \gamma}^{\tmop{SM}}$, i.e $C_{7 \gamma}
\simeq |C_{7 \gamma}^{\tmop{SM}} | e^{i \theta}$.  Let us take an
illustrative example in which the new physics contribution is
purely electro(chromo)-magnetic and satisfy  $C_{7 \gamma}^{} =
|C_{7 \gamma}^{\tmop{SM}} | e^{i \theta}$ and also $C_{8 g}^{} =
|C_{8 g}^{\tmop{SM}} | e^{i \theta}$ at the scale of $m_W$.
Varying the value  of $\theta$ from 0 to $2 \pi$ and then running
down to the low energy scale of $\mu \simeq m_b$ through
renormalization group equation, one finds that the value of $\sin
2 \beta_{\tmop{eff}}$ in decay $B \rightarrow \phi K_S$  only
changes from 0.5 to 0.8, This naive discussion shows that  if the
dominant contribution from a new physics model is coming from
$C_{7 \gamma ( 8 g )}^{\tmop{NEW}}$, the change to  $\sin 2
\beta_{\tmop{eff}}$ from the its SM value is limited. Unfortunately, the
S2HDM belongs to this class of model. The recent analysis have
confirmed that within S2HDM, the value of $\sin 2
\beta_{\tmop{eff}}$ can reach zero, but not likely to be largely
negative\cite{Hiller:2002ci,Giri:2003jj,Huang:2003bj}.
 
For the model of SM4, there are additional up ($t'$) and down ($b'$) type
quarks. The new phases may come from the  extended
Cabbibo-Kobayashi-Maskawa(CKM) matrix which is a four by four matrix in this
model and contains undetermined matrix elements of $V^{}_{t' q}, V_{q b'}$
etc. To avoid the precise data of electro-weak processes, the mass of $b'$ (
$t'$ ) has to be pushed to greater than $\sim$200 GeV( $\sim 300$ GeV). However,
phenomenological study showed that with the constraint of $B \rightarrow X_s
\gamma$  and $B^0_s - \overline{B^0}_s^{}$ mixings being considered, its
contribution to the CP violation of $B \rightarrow \phi K_S$ is not large
enough either\cite{Arhrib:2002md}.  Thus if the large negative value of $\sin 2 \beta_{eff}$ in decay $B
\rightarrow \phi K_S$ is confirmed by the future experiments, the above
mentioned two models ( i.e. S2HDM and SM4 ) will not be favored.
 
section{The model of S2HDM4}
 
There are several directions in constructing models beyond the SM, such as
enlarging the gauge groups to $\tmop{SU} ( 5 )$, $\tmop{SU} ( 10 ) \tmop{and}
E_6$ etc.,  introducing new symmetries like various SUSY models, and expanding the
matter contents, i.e., more fermions and Higgs bosons. The models of
the last type  can be regarded as simple  extensions of the SM which keep the
same gauge structure but still have rich sources of new  contributions. The
typical ones are the above mentioned S2HDM and SM4.
 
In this paper we would like to a step further to consider a model with both two-Higgs-doublet and
fourth-generation quarks (S2HDM4). In this model, there are new Yukawa
interactions between Higgs bosons and heavy fourth-generations quarks. Since
in general the Yukawa interaction is expected to be proportional to the coupled quark mass,
 the new Yukawa couplings are  much  stronger than that in the S2HDM and SM4 . Unlike in the case of S2HDM,
where the $b$ quark contribution to the QCD penguin diagram through neutral
Higgs boson loop  is  strongly suppressed by the small $b$ quark mass, the
same diagram with intermediate $b'$ quark may significantly contribute to the
related processes \cite{Wu:1998ng}. This new feature only exists  in this combined model,  and is
of particular interest in studying the CP violation  of $B \rightarrow \phi K_S $  and other
penguin dominant processes.
 
The Lagrangian for the S2HDM4  is given by
\begin{eqnarray}
  \mathcal{L}_Y & = & \bar{\psi}_L Y^U_1  \widetilde{\phi_1} u_R +
  \bar{\psi}_L Y^D_1 \phi_1 d_R + \bar{\psi}_L Y^U_2  \widetilde{\phi_2} u_R +
  \bar{\psi}_L Y^D_2 \phi_2 d_R + H.c
\end{eqnarray}
with the extended quark content of  $u_{L, R} = ( u, c, t, t' )_{L, R}$ and
$d_{L, R} = ( d, s, b, b' )_{L, R}$. The Yukawa coupling matrices $Y^{U ( D
)}_i$ are 4-dimensional matrices accordingly. The two Higgs fields $\phi_1, \phi_2$ have vacuum
expectation values (VEV) of $v_1 e^{i \delta_1}$ and $v_2 e^{i \delta_2}$ respectively,
with $\sqrt{|v_1 |^2 + |v_1 |^2} = v = 246 \tmop{GeV} .$ The relative phase
$\delta = \delta_1 - \delta_2$ between two VEVs is physical and provides a new
source of CP violation\cite{wu:1994ja,wolfenstein:1994jw,Wu:1994vx} .  In the mass eigenstates, the three physical Higgs
bosons are denoted by $H^0, A^0, \tmop{and} H^{\pm}$ respectively.  Due to the non-zero
phase $\delta$,  all the Yukawa couplings become complex numbers in the
physical mass basis, even they are all real in the flavor basis. For simplicity,
throughout this paper, we assume that the CKM matrix elements associating with $t'$, i.e.  $V_{t'q}$
are ignorablly small  and will only focus on the neural Higgs boson contributions.
 
In the mass basis, the Yukawa interactions between neutral Higgs bosons  and
quarks   have the following general form
\begin{eqnarray}
  \mathcal{L}_Y & = & \eta^q_{ij}  \bar{q}_{iL} q_{jR} \phi + H.c.,
\end{eqnarray}
with $\phi = H^0$ or $A^0$. The Yukawa coupling $\eta^q_{\tmop{ij}}$ is
usually parameterized as
\begin{eqnarray}
  \eta_{\tmop{ij}}^q & = &  \frac{\sqrt{m_{q_i} m_{q_j}}}{v} \xi_{q_i
  q_j}
\end{eqnarray}
In the Chen-Sher ansartz {\cite{cheng:1987rs}} motivated by a Fritzsch type of
Yukawa coupling matrix. the values of all $\xi_{q_i q_j}$s are of the same
order of magnitude. However, from other textures of the coupling matrix the relations among
$\xi_{q_i q_j}$s are different\cite{Zhou:2003kd,Diaz-Cruz:2004tr,Diaz:2003za}. In the general case, they should be taken  as
free parameters to be determined or constrained by the experiments.
 
The effective Hamiltonian for $\Delta B = 1$ charmless $B$ decays reads
\begin{eqnarray}
  H_{\tmop{eff}} & = & \frac{G_F}{\sqrt{2}} \left[ V_{\tmop{ub}}
  V_{\tmop{us}}^{*} ( C_1^u Q_1^u + C_2^u Q_2^u ) + V_{\tmop{cb}}
  V_{\tmop{cs}}^{*} ( C_1^c Q_1^c + C_2^c Q_2^c ) \right. \nonumber\\
  &  & \left. - V_{\tmop{tb}} V_{\tmop{ts}}^{*} \left( \sum_{i = 3}^{10}
  C_i Q_i + C_{7 \gamma} Q_{7 \gamma} + C_{8 g} Q_{8 g} \right) \right],
\end{eqnarray}
where the operator basis  $Q_i$s  can be found in Ref.\cite{Buras:1998ra}. In this model, the relevant
Wilson coefficients at the scale of  $m_W$ from this model is given by
\begin{eqnarray}
  C_1 ( M_W ) & = & \frac{11}{2}  \frac{\alpha_s ( M_W )}{4 \pi} ,\nonumber\\
  C_2 ( M_W ) & = & 1 - \frac{11}{6}  \frac{\alpha_s ( M_W )}{4 \pi} -
  \frac{35}{18}  \frac{\alpha_{\tmop{em}}}{4 \pi} ,\nonumber\\
  C_3 ( M_W ) & = & - \frac{\alpha_s ( M_W )}{24 \pi} ( \tilde{E}_0 ( x_t ) +
  | \xi_{t t} |^2 E_0^{\tmop{III}} ( y ) + \frac{m_{b'} \sqrt{m_b m_s}}{2 V_{t
  b}^{} V_{t s}^{*} m_t^2} \xi_{b b'}^{*} \xi_{s b'}
  E_0^{\tmop{III}} ( y' ) ) ,\nonumber\\
  &  & + \frac{\alpha_{\tmop{em}}}{6 \pi}  \frac{1}{\sin^2 \theta_W} ( 2 B_0
  ( x_t ) + C_0 ( x_t ) ) ,\nonumber\\
  C_4 ( M_W ) & = & \frac{\alpha_s ( M_W )}{8 \pi} ( \tilde{E}_0 ( x_t ) + |
  \xi_{t t} |^2 E_0^{\tmop{III}} ( y ) + \frac{m_{b'} \sqrt{m_b m_s}}{2 V_{t
  b}^{} V_{t s}^{*} m_t^2} \xi_{b b'}^{*} \xi_{s b'}
  E_0^{\tmop{III}} ( y' ) ) ,\nonumber\\
  C_5 ( M_W ) & = & - \frac{\alpha_s ( M_W )}{24 \pi} ( \tilde{E}_0 ( x_t ) +
  | \xi_{t t} |^2 E_0^{\tmop{III}} ( y ) + \frac{m_{b'} \sqrt{m_b m_s}}{2 V_{t
  b}^{} V_{t s}^{*} m_t^2} \xi_{b b'}^{*} \xi_{s b'}
  E_0^{\tmop{III}} ( y' ) ) ,\nonumber\\
  C_6 ( M_W ) & = & \frac{\alpha_s ( M_W )}{8 \pi} ( \tilde{E}_0 ( x_t ) + |
  \xi_{t t} |^2 E_0^{\tmop{III}} ( y ) + \frac{m_{b'} \sqrt{m_b m_s}}{2 V_{t
  b}^{} V_{t s}^{*} m_t^2} \xi_{b b'}^{*} \xi_{s b'}
  E_0^{\tmop{III}} ( y' ) ) ,\nonumber\\
  C_{7 \gamma} ( M_W ) & = & \frac{A ( x_t )}{2} - \frac{1}{2} \left( A ( y )
  | \xi_t |^2 + A ( y' ) \frac{m_{b'} \sqrt{m_b m_s}}{2 V_{t b}^{} V_{t
  s}^{*} m_t^2} \xi_{b b'}^{*} \xi_{s b'} \right) ,\nonumber\\
  &  & + B ( y ) | \xi_t \xi_b | e^{i \theta} - B ( y' ) \frac{m_{b'}
  \sqrt{m_b m_s}}{2 V_{t b}^{} V_{t s}^{*} m_t^{} m_b^{}} \xi_{b' b}^{}
  \xi_{s b'} ,\nonumber\\
  C_{8 g} ( M_W ) & = & - \frac{D ( x_t )}{2} - \frac{1}{2} \left( D ( y ) |
  \xi_t |^2 + D ( y' ) \frac{m_{b'} \sqrt{m_b m_s}}{2 V_{t b}^{} V_{t
  s}^{*} m_t^2} \xi_{b b'}^{*} \xi_{s b'} \right) ,\nonumber\\
  &  & + ( y ) | \xi_t \xi_b | e^{i \theta} - E ( y' ) \frac{m_{b'} \sqrt{m_b
  m_s}}{2 V_{t b}^{} V_{t s}^{*} m_t^{} m_b^{}} \xi_{b' b}^{} \xi_{s
  b'}^{*},
\end{eqnarray}
with $\alpha_s ( m_W )$ and $\alpha_{\tmop{em}}$ being the strong
and electro-magnetic couplings at scale $m_W$. The mass ratios
$x_t, y$ and $y'$ are defined as $x_t = m_t^2 / m_W^2$, $y = m_t^2
/ m_{H^{\pm}}^2$ and  $y' = m_{b'}^2 / m_{H^0}^2$ respectively.
The loop integration functions are standard and can be found in
Refs.\cite{Inami:1981fz,Gilman:1980di,Gilman:1983ap,Xiao:2002mr}.
Here we have ignored the coefficients for the electro-weak penguin
diagrams since their effects are less significant in the decay of
$B \rightarrow \phi K_S$.
 
Note that the new contributions to QCD  and
electro(chromo)-magnetic operators depends on different parameter
sets. In the QCD penguin sector, the contribution depends on
$\xi_{b b'}^{*} \xi_{s b'}$ where in electro(chromo)-magnetic
sector it depends on both $\xi_{b' b}^{} \xi_{s b'}$ and
$\xi_{b b'}^{*} \xi_{s b'}$. It is convenient to define two
weak phases $\theta_1$ and $\theta_2 $ with
\begin{eqnarray}
  &  & \xi_{b b'}^{*} \xi_{s b'} = | \xi_{b b'} \xi_{s b'
  } | e^{i \theta_1} \quad \mbox{and}\quad \xi_{b' b}^{} \xi_{s b'} = |
  \xi_{b' b}^{} \xi_{s b' } |e^{i \theta_2} .
\end{eqnarray}
Since in general $\xi_{b' b}^{}$ and $\xi_{b b'}^{*} $ are complex numbers
and $\xi_{b' b}^{} \neq \text{$\xi_{b b'}^{*} $}$, the two phases are not
necessary to be equivalent. The presence of two rather than one independent
phases is particular for this model, which gives different contributions to the
QCD penguin and electro(chromo)-magnetic Wilson coefficients.  The
interference between them enlarges the allowed parameter space.
Note that the Wilson coefficient for QCD penguins may be complex
numbers which  provides additional sources of CP violation. To make
a comparison, let us denote the Wilson coefficients in the SM by
$C^{SM}_i$. Taking $\xi_{b'b}=\xi_{s b'}=0.8$, $\theta_{1}=0.5$,
$\theta_2=-1.2$ and $m_{H^0}=m_{b'}=200$GeV as an example, in the
range of $40<\xi_{bb'}<60$, the ratio of
$C_{3}/C_{3}^{SM}(C_{4}/C_{4}^{SM})$ has an imaginary part between
-0.27 and -0.4(-0.6 and -0.8). These large imaginary parts plays
an important role in  CP violation.

\section{Constraints from $B \rightarrow X_s \gamma$ and$B^0_s - \bar{B}^0_s$
mixing}
 
Before making any predictions, one first needs to know how the new
parameters in this model are constrained by other experiments. For
the process we are concerning, the most strict constraints comes
from $b \rightarrow s \bar{s} s$ processes such as $B \rightarrow
X_s \gamma$ and $B^0_s - \bar{B}^0_s$ mixing, etc.
 
\begin{figure}[htb]
\includegraphics[width=0.8\textwidth]{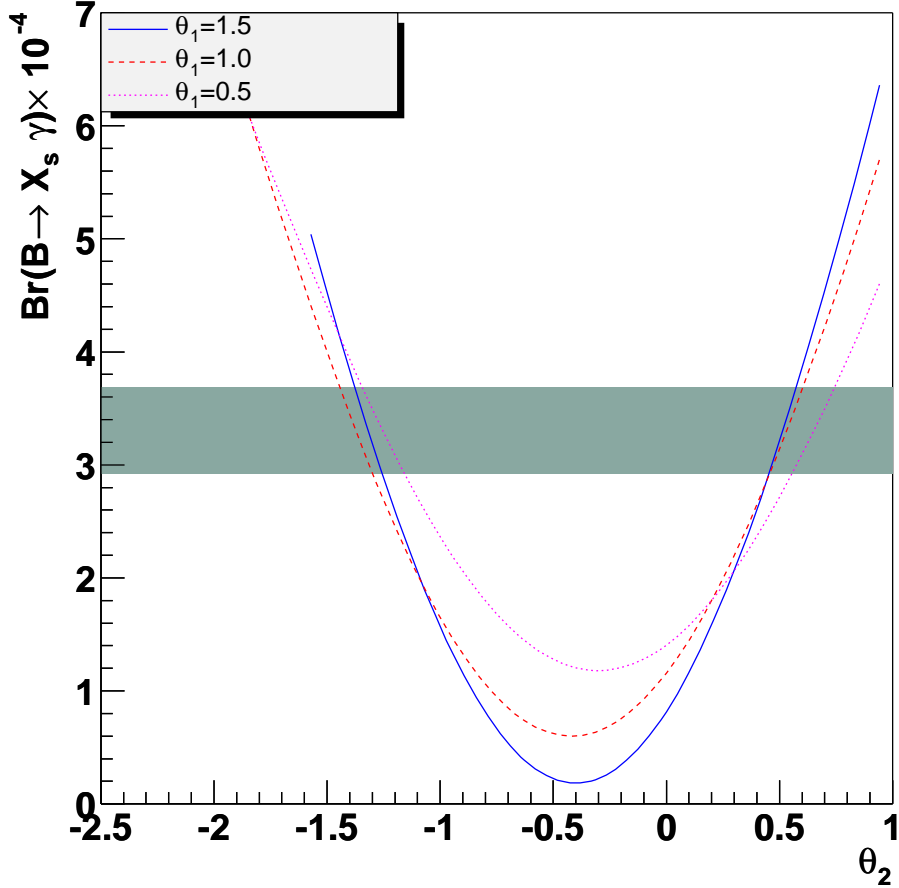}
\caption{The branching
ratio of $B \rightarrow X_s \gamma$ as a functions of $\theta_2$ in the model of
S2HDM4 . The solid, dashed and dotted curves correspond to $\theta_1 = 1.5, 1.0
\tmop{and} 0.5$ respectively. Other parameters are taken from
Eq.(\ref{params}).}
\label{BSG}
\end{figure}

The expression for $B \rightarrow X_s \gamma$ normalized to $B \rightarrow
X_c e \bar{\nu}_e$ reads
\begin{eqnarray}
  \frac{\tmop{Br} ( B \rightarrow X_s \gamma )}{\tmop{Br} ( B \rightarrow X_c
  e \bar{\nu}_e )} & = & \frac{6 |V_{\tmop{tb}} V_{\tmop{ts}}^{*} |^2
  \alpha_{\tmop{em}}}{\pi |V_{\tmop{cb}} |^2 f ( m_c / m_b )} |C_{7 \gamma} (
  \mu ) |^2
\end{eqnarray}
with $f ( z ) = 1 - 8 z^2 - 24 z^4 \ln z + 8 z^6 - z^8$ and
$\tmop{Br} ( B \rightarrow X_c e \bar{\nu}_e ) = 10.45\%$. The low
energy scale $\mu$ is set to be $ m_b$. Using the Wilson
coefficients at the  scale $m_W$ and running down to the $m_b$ scale
through re-normalization group equations, we obtain the
predictions for  Br($B \rightarrow X_s \gamma$).  For simplicity, we
focus on the case in which the $b'$ contribution  dominates
through $H^0$ loop, namely, we push the masses of the charged
Higgs $H^{\pm}$ and the other pseudo-scalar boson $A^0$  to be very high
$( m_{H^{\pm}}, m_{A^0} > 500$ GeV) and ignore their
contributions. We take the following typical values of the
couplings
\begin{equation}
  | \xi_{b b'} | = 50, \quad | \xi_{b' b} | = 0.8, \quad | \xi_{s b'} | = 0.8,  \quad \tmop{and}
  \quad m_{H^0} = m_b' = 200 \tmop{GeV}, \label{params}
\end{equation}
and give in Fig.\ref{BSG} the value of Br($B \rightarrow X_s \gamma$) as a
function of $\theta_1$ with different values of $\theta_2$.
 
 
\begin{figure}[htb]
\includegraphics[width=0.8\textwidth]{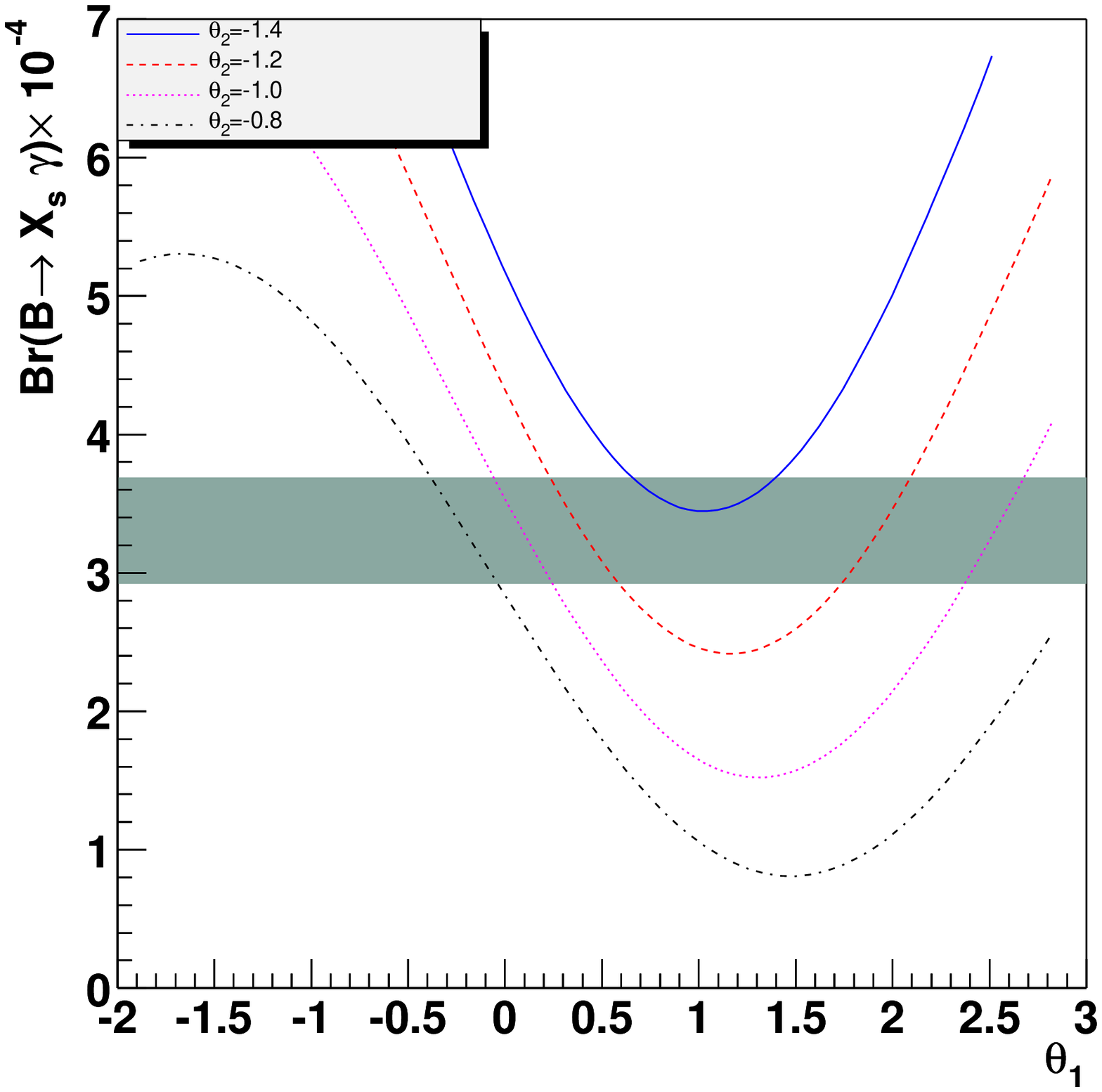}
\caption{The branching ratio of $B \rightarrow X_s \gamma$ as a
functions of $\theta_1$ in the model of S2HDM4 . The solid, dashed, dotted  and
dot-dashed curves correspond to $\theta_2 = 1.4, 1.2, 1.0 \tmop{and} 0.8$
respectively. Other parameters are taken from Eq.(\ref{params})}.
\label{BSG-case2}
\end{figure}

From the figure, one finds that two separated  ranges for
parameters $\theta_1$ and $\theta_2$ are allowed by the data
\begin{eqnarray}
  &  & \text{$- 1.4 \lesssim \theta_2 \lesssim - 1.2$} \tmop{and} \text{ $0.4
  \lesssim \theta_2 \lesssim 0.7$} \qquad \tmop{for} \text{$0.5 \lesssim \theta_1
  \lesssim 1.5$},
\end{eqnarray}
Note that we do not make a scan for the full parameter space, nevertheless the
above obtained range are already enough for our purpose. Among the two allowed
ranges, the one with $- 1.4 \lesssim \theta_2 \lesssim - 1.2$ is of particular
interest. It will be seen below that in this range, the contribution to the CP
asymmetry in $B \rightarrow \phi K_S$ could be significant. In
Fig.\ref{BSG-case2}, we also give the allowed range of $\theta_1$ with
difference values of $\theta_2$.  One finds that the allowed range for
$\theta_1$ is larger compared with $\theta_2$. In this figure, the interference
between two phases $\theta_1$ and $\theta_2$ is manifest. For $\theta_2$ in the
range of $( - 1.0, - 0.8 )$, the allowed value for $\theta_1$ is a narrow window
around zero. But for $\theta_2$ in the range of $( - 1.4, - 1.2 )$, the allowed
range for $\theta_1$ could be between 0.5 and 2.0. Compared with the S2HDM in
which  only one phase appears, this interference effect for two phases
enlarges the parameter space under the constraint of $B \rightarrow X_s \gamma$.
Thus large contributions to the other processes is possible in this model.
 
%
 
The other $b \rightarrow s \bar{s} s$ process which could impose
strong constraint is the mass difference of neutral $B_s^0 $
meson. The measurements from LEP give a lower bound of $\Delta
m_{B_s}>14.9 ps^{-1}$. In this model, the $b'$ contributes to $\Delta
m_{B_s}$ only through box-diagrams. The box diagram contribution to
$\Delta m_{B_s}$ is given by
{\cite{Inami:1981fz,Gilman:1980di,Gilman:1983ap,wu:1999fe}}
\begin{eqnarray}
  \Delta m_{B_s} & = & \frac{G_F^2}{6 \pi^2} ( f_{B_s} \sqrt{B_{B_s}} )^2
  m_{B_s} m_t^2 |V_{\tmop{ts}} |^2 \{ \eta_{\tmop{tt}} B^{\tmop{WW}} ( x_t ) +
  \frac{1}{4} \eta^{\tmop{HH}}_{\tmop{tt}} y_t | \xi_{tt} |^4 B^{\tmop{HH}}_V (
  y_t ) \nonumber\\
  &  & + 2 \eta_{\tmop{tt}}^{\tmop{HW}} y_t | \xi_{\tmop{tt}} |^2
  B^{\tmop{HW}}_V ( y_t, y_w ) + \frac{1}{4} \eta^{\tmop{HH}}_{\tmop{tt}} y' (
  \frac{m_{b'} \sqrt{m_b m_s}}{2 V_{t b}^{} V_{t s}^{*} m_t^2} \xi_{b
  b'}^{*} \xi_{s b'}^{} )^2 B^{\tmop{HH}}_V ( y' ) \}
\end{eqnarray}
where $G_F = 1.16 \times 10^{- 5} \tmop{GeV}^{- 2}$ is the Fermi constant.
$f_{B_s}$ and $B_{B_s}$ are the decay constant and bag parameter for $B_s^0$. In the numerical
calculations,  we take the value of $f_{B_s} \sqrt{B_{B_s}} =0.23$GeV.
$\eta_{i j}$s are the QCD correction factors. The loop integration functions
of $B^{\tmop{HH}, \tmop{WW}, \tmop{HW}}_{(V)}$ can be found in Refs\cite{Gilman:1980di,Gilman:1983ap,wu:1999fe}.
The mass ratios
are defined as $y_t = m_t^2 / m_{H^{\pm}}^2, y_w = m_t^2 / m_W^2$ and $y' =
m_{b'}^2 / m_{H^0}^2$ respectively.  Note that in the mass difference of $B^0_s$ mesons,
the contribution from S2HDM4 only depends on the parameter $\xi_{b
b'}^{*} \xi_{s b'}$. So, only the phase $\theta_1$ will present in the
expression.
 
\begin{figure}[htb]
\includegraphics[width=0.8\textwidth]{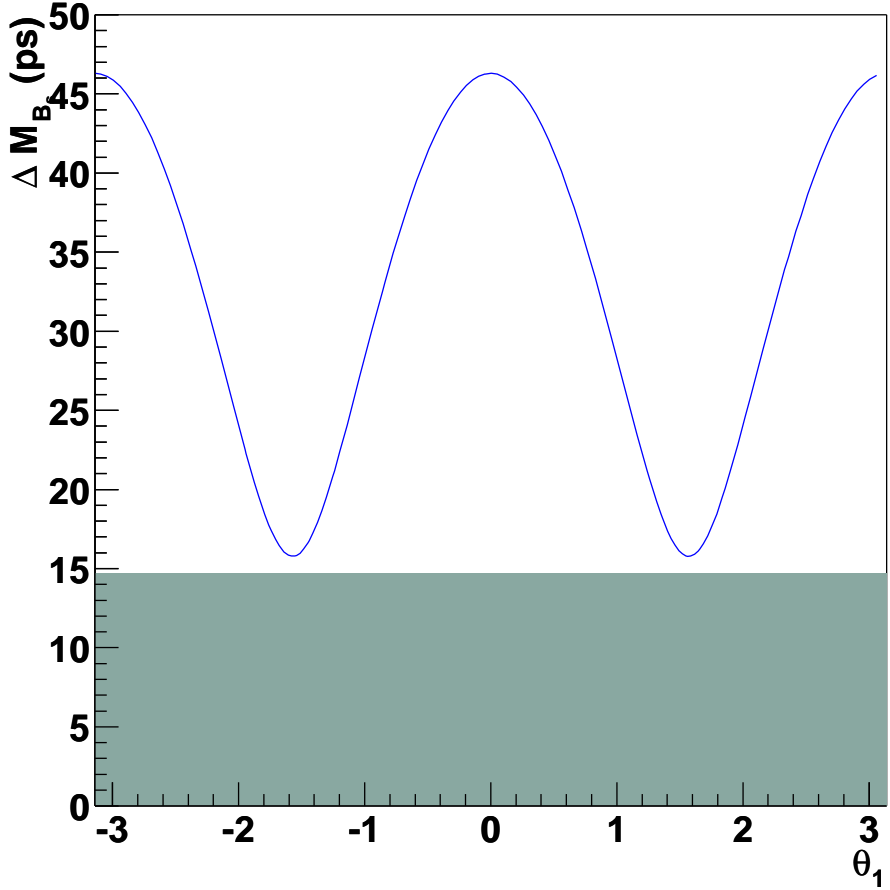}
\caption{The $B^0_s$ meson
mass difference $\Delta B^0_s$ as a function of $\theta_1$ in the model of
S2HDM4 .  Other parameters are taken from Eq.(\ref{params}). The shadowed
region is excluded by the data of $\Delta B^0_s$.}
\label{B0s}
\end{figure}

Using the above obtained typical parameters in Eq.(\ref{params}), the contribution to $\Delta
m_{B_s}$ is  calculated and  plotted as a function of $\theta_1$ in
Fig.\ref{B0s}.  The figure shows that the current data of $\Delta m_{B_s}$ do
not impose strong constraint on the value of $\theta_1$.
 
The neutron electric dipole moment (EDM) is expected to give strong constraints
on the new physics. In the SM, the neutron EDM is zero at even two loop
level. The current experimental upper limit gives EDM$<1.1\times 10^{-25}
ecm.$\cite{Hagiwara:2002fs}.  In general, the new physics contributes to the neutron EDM
through one loop diagrams. In the presence of new scalars, additional
significant contributions may arise, for example from the Weinberg gluonic operator
\cite{Weinberg:1989dx} and also the two-loop Barr-Zee type diagrams \cite{Bjorken:1977vt,Barr:1990vd} 
etc.

However, we note that all the above three type of mechenisms are not related to
$b\to s$ flavor-changing transitions and therefore will involve different
parameters in this model. For the one-loop diagrams, the neutral EDM is mostly
related to $\xi_{u(d)}$ and $\xi_{t(b')}$ through $u(d)-$quark EDM. For Weinberg
three gluonic operator, the dominate contribution is from intereral $b'$
loop. Thus it is related to $\xi_{b'b'}$.  Similarly, for two-loop Barr-Zee
diagram, the $b'-$quark loop will play the most important role and the couplings
involve only $\xi_{u(d)}, \xi_{b'b'}$ etc.
Thus the neutron EDM will impose strong constraints on other paramerters in this
model and has less significance in current studying of decay $B\to \phi K_S$.
This is significantly different from the S2HDM case in which the $t-$quark alway
domains the loop contribution and the couplings $\xi_{tt}$ and $\xi_{bb}$ are
subjected to a strong constraint from neutron EDM.

Other constraints may  come from $K^0 - \overline{K^0}$ and $B^0_d -
\overline{B^0_d} $ mixings. But those processes contain additional free
parameters such as the the Yukawa coupling of $\xi_{b' d}$ and $\xi_{s b'}$,
the constraints from those processes are much weaker.
 
 
\section{CP asymmetry in $B \rightarrow \phi K_S$}
 
Now we are in the position to discuss CP asymmetry in $B \rightarrow \phi
K_S$. The decay amplitude for $\bar{B} \rightarrow \phi \bar{K^0}$reads
 
\begin{eqnarray}
  \mathcal{A}( \bar{B}^0_d \rightarrow \phi \bar{K}^0 ) & = & -
  \frac{G_F}{\sqrt{2}} V_{\tmop{ts}}^{*} V_{\tmop{tb}} ( a_3 + a_4 + a_5 -
  \frac{1}{2} ( a_7 + a_9 + a_{10} ) ) X,
\end{eqnarray}
with $X$ being a factor related to the hadronic matrix elements. In the naive
factorization approach $X = 2 f_{\phi} m_{\phi} ( \epsilon \cdot p_B ) F_1 (
m_{\phi} )$, where $\epsilon$, $p_B$, $F_1$ are the polarization vector of $\phi$,
the momentum of $B$ meson and form factor respectively.
The coefficients $a_i $ are defined through the effective Wilson
coefficients $C^{\tmop{eff}}_i s$ as follows
\begin{eqnarray}
  a_{2 i - 1} = C^{\tmop{eff}}_{2 i - 1} + \frac{1}{N_c} C^{\tmop{eff}}_{2 i}
  & , & a_{2 i} = C^{\tmop{eff}}_{2 i} + \frac{1}{N_c} C^{\tmop{eff}}_{2 i -
  1},
\end{eqnarray}
Since the heavy particles such as $H^{\pm, 0}, A^0 \tmop{and} b'$ has
been integrated out below the scale of $m_W$, the procedures to obtain
the effective Wilson coefficients $C^{eff}_i$ are exactly the same as in SM
and can be found in Ref.\cite{Buchalla:1996vs} .
 
\begin{figure}[htb]
\includegraphics[width=0.8\textwidth]{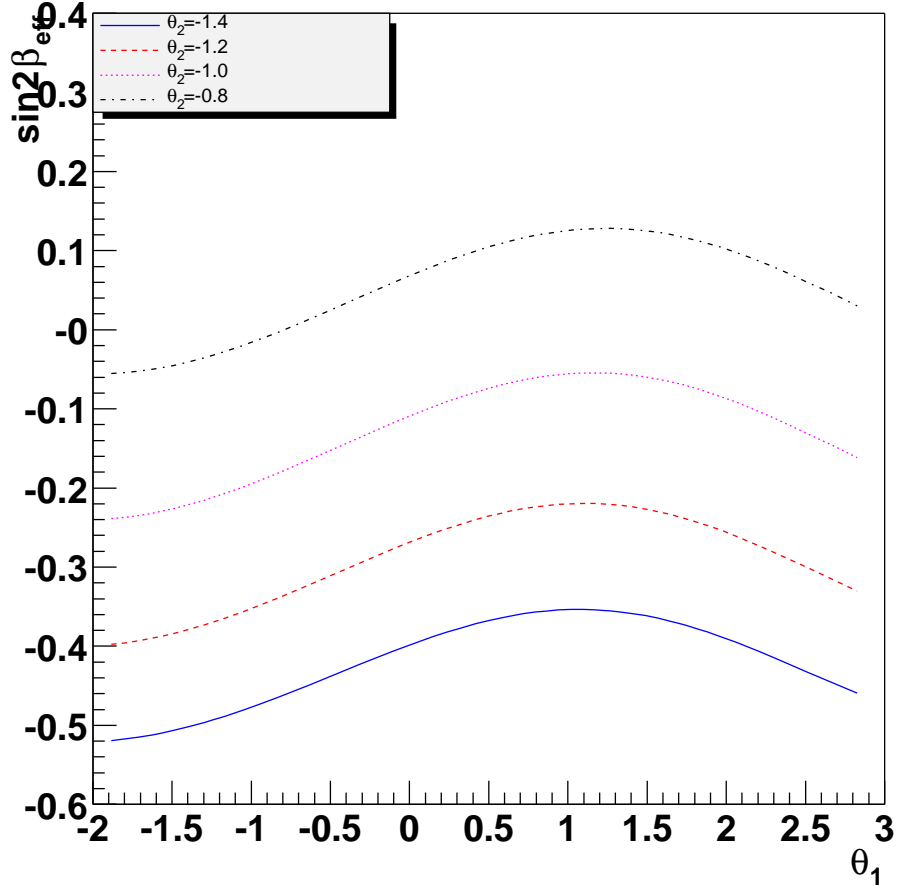}
\caption{The prediction for $\sin 2
\beta_{\tmop{eff}}$ as a functions of $\theta_1$ with different value of
$\theta_2$. The solid, dashed, dotted and dot-dashed curves corresponds to
$\theta_2 = - 1.4, - 1.2, - 1.0, - 0.8$ respectively.}
\label{BPHIK}                   %
\end{figure}

Using the above obtained parameters allowed by the current data, the prediction for
the time dependent CP asymmetry for $B \rightarrow \phi K_S$  are shown in Fig.\ref{BPHIK}
 
%
 
In the figure, we give the value of $\sin 2 \beta_{\tmop{eff}}$ as
a function of $\theta_1$ with different values of
$\theta_2$=1.4,1.2,1.0 and 0.8. Comparing with the constraints
obtained from $B \rightarrow X_s \gamma$ and $B^0_s-\bar{B}^0_s$
mixings, one sees that in the allowed range of  $\text{$- 1.4 <
\theta_2 < - 1.2$} $ and $0.5 < \theta_1 < 1.5$, the predicted
$\sin 2 \beta_{\tmop{eff}}$ can reach $- 0.4$.
 
It is evident that the large negative value of $\sin2\beta_{eff}$ is a
consequence of the interference effects between $\theta_1$ and $\theta_2$ and
therefore is particular for this model. For zero value of $\theta_1$, there is
no new phase in the QCD penguin sector. From Fig.\ref{BSG-case2}, the
allowed range for $\theta_2$ is $-1.0 \alt \theta_2 \alt -0.8$. Then, it follows from
Fig.\ref{BPHIK}, that in this range the predicted $\sin2\beta_{eff}$
is at around zero.  But for $\theta_1 \approx 0.5$, the allowed range for $\theta_2$
is changed into  $-1.4 \alt \theta_2 \alt -1.2$ and the predictions for $\sin2\beta_{eff}$ %
is much lower in the range of $(-0.4,-0.25)$.

\section{Conclusions}
 
In conclusion, we have discussed the CP asymmetry of  decay $B
\rightarrow \phi K_S$, in the model of S2HDM4 which contains both
an additional Higgs doublet and fourth generation quarks. In this
model, since the fourth generation $b'$ quark is much heavier that
$b$ quark, the Yukawa interactions between neutral Higgs boson and
$b'$ is greatly enhanced. This  results in significant modification
to the QCD penguin diagrams. We have obtained the allowed range of
the parameters  from the process of $B \rightarrow X_s \gamma$ and
$\Delta m_{B_s^{}}$. Due to the more complicated phase effects, in
this model the constraints from those process are weaker than that
in S2HDM and SM4. The
 effective $\sin 2 \beta_{\tmop{eff}}$ in the decay $B \rightarrow \phi K_S$ is
predicted with  the constrained parameters. We have found that
this model can easily account for the possible large negative
value of $\sin 2 \beta$ without conflicting with other
experimental constraints.

In this paper we focus on the case in which $H^0$ domains. It is straight
forward to find that the contribution from the other pseudo-scalar $A^0$ follows
the same pattern. In the case of small mixing among the neutral scalars, the
Yukawa couplings for $H^0$ and $A^0$ are directly related \cite{atwood:1997vj}. We find
that for $m_{A^0}\approx 200 \mbox{ GeV } \ll m_{H^0}$ its contribution to the decay amplitude of $B \to \phi K_S$ is similar to the case of the $H^0$ 
dominance discussed above. For the case that $m_{A^0}$ is close
to $m_{H^0}$, the contribution from them are comparable, and the interference between the two could be important.

Since this model contributes new phases to QCD penguin diagrams,
it remains to be seen if it has sizable effects on other penguin
dominant processes, such as in the hadronic charmless B decays.
Similarly, it is expected that in this model there are also
significant contributions to the electro-weak (EW) penguin diagrams
which deserves a further investigation 
(for recent discussions on EW penguin effects on $B\to \phi K$ see, 
e.g\cite{Atwood:2003tg,Morrissey:2003sc,Deshpande:2003nx}.)
It is well known that the EW penguin plays important roles in rare B decays.
The current data on $B\rightarrow\pi\pi, \pi K$ have indicated
some deviations from results based on the SM
\cite{Aubert:2003hf,Abe:2003yy,
Abe:2004us,Zhou:2000hg,Wu:2002nz,Buras:2003dj}. It is of interest
to further investigate the new physics contributions to those
decay modes within this model.

\begin{acknowledgments}
YLW was supported in part by the key projects
of Chinese Academy of Sciences and National Science Foundation of China
(NSFC).
\end{acknowledgments}
 
 

\bibliographystyle{apsrev} \bibliography{/home/zylon/zhou/reflist/reflist.bib}
 
\end{document}